\documentclass[aps,prx,reprint,showpacs,twocolum]{revtex4-2}
\usepackage{amsmath,amssymb,graphicx}
\usepackage{amsmath,amssymb,graphicx}
\usepackage{multirow}
\usepackage{booktabs}
\usepackage{makecell}
\usepackage{color}
\usepackage{float}
\usepackage{titletoc}
\usepackage[singlelinecheck=false,justification=raggedright]{caption}
\usepackage{subcaption}
\newcommand{\bitem}{\begin{itemize}}
	\newcommand{\fitem}{\end{itemize}}
\newcommand{\beq}{\begin{equation}}
	\newcommand{\eeq}{\end{equation}}
\newcommand{\beqa}{\begin{eqnarray}}
	\newcommand{\eeqa}{\end{eqnarray}}

\providecommand{\U}[1]{\protect\rule{.1in}{.1in}}
\begin{document}
\title{A Squeezed Vacuum State Laser with Zero Diffusion}
\author{F. de Oliveira Neto$^{1}$, G. D. de Moraes Neto$^{2,*}$, and M. H. Y.
Moussa$^{1}$}
\affiliation{$^{1}$Instituto de F\'{\i}sica de S\~{a}o Carlos, Universidade de S\~{a}o
Paulo, Caixa Postal 369, 13560-970, S\~{a}o Carlos, S\~{a}o Paulo, Brazil}
\affiliation{$^{2}$Department of Physics, Zhejiang Normal University, Jinhua 321004,
People's Republic of China }
\email{gdmneto@gmail.com}
\begin{abstract}
We propose a method for building a squeezed vacuum state laser with zero
diffusion, which results from the introduction of the reservoir engineering
technique into the laser theory. As well as the reservoir engineering, our
squeezed vacuum laser demands the construction of an effective atom-field
interaction. And by building an isomorphism between the cavity field operators
in the effective and the Jaynes-Cummings Hamiltonians, we derive the equations
of our \textit{effective laser}\ directly from the conventional laser theory.
Our method, which is less susceptible to errors than reservoir engineering,
can be extended for the construction of other nonclassical state lasers, and
our squeezed vacuum laser can contribute to the newly emerging field of
gravitational interferometry.

\end{abstract}

\maketitle

\section{Introduction}

The laser theory is one of the central developments in the physics of
radiation-matter interaction. Based on the theoretical framework provided by
Townes and Schawlow \cite{TS}, the first laser, built by Maiman \cite{Maiman},
date back to the 1960s, and has since played a major role both in basic and
applied physics, with applications in many technical aspects of modern
society. The quantum theory of the laser was built basically from
contributions led by H. Haken \cite{Haken}, W. E. Lamb \cite{Lamb}, and M. Lax
\cite{Lax}, from which are derived the more realistic models where a
transmitting window \cite{Baseia}, and the pumping statistics of the lasing
atoms \cite{RW} are included.

Among many others, we mention the uses of lasers for cooling and trapping
atoms \cite{1997}, for Bose-Einstein condensation in dilute gases of alkali
atoms \cite{2001}, for the development of optical tweezers and their
application in biological and physical sciences \cite{Tweezers}, and for
generating ultrashort high-intensity laser pulses extensively used across
physics and chemistry \cite{UHILP}. These achievements draw a broader picture
of the unique progress that quantum optics has undergone since the 1980s. In
addition to this picture we mention the generations of squeezed states of the
radiation field \cite{SS}, essential for enhancing interferometric sensitivity
\cite{IS}, and today a critical challenge for the development of gravitational
wave interferometry \cite{Caves,Meystre,Ligo}. We also mention the
applications of squeezed states in optical waveguide tap \cite{Shapiro},
quantum nondemolition measurements \cite{Meystre}, quantum information
processing \cite{Furusawa}, and quantum metrology \cite{Metrology}. There is
also the development of different sources for generating entangled photon
states, used for investigating fundamentals of quantum mechanics
\cite{Photons}.

Parallel to the developments of quantum optics, we witnessed the emergence of
quantum communication and computation, which resulted in the new and promising
field of quantum information theory \cite{TIQ}. The need for implementation of
quantum logic operations demanded new techniques for engineering nonclassical
states \cite{StateEng}, effective interactions \cite{DJ,EH} and reservoirs
\cite{PCZ,ER} for phase coherence control. These demands have pushed the
physics of the radiation-matter interaction to a new level through platforms
such as cavity quantum electrodynamics \cite{Haroche}, trapped ions
\cite{Wineland}, circuit quantum electrodynamics \cite{CQED}, and all related topics.
Regarding coherence control, a key issue for the construction of a squeezed
vacuum laser with zero diffusion, many methods have been designed
\cite{Methods}; however, the reservoir engineering \cite{PCZ} ---which seems
inspired by the laser theory--- is of particular interest here. Its basic idea
is to submit the system of interest, say a dissipative cavity mode (described
by the creation and annihilation operators $a^{\dag}$ and $a$, and whose
particular state $\left\vert \Psi\right\rangle $ we intend to protect from the
action of the environment), to an interaction with an auxiliary strongly
dissipative system as, for example, a two-level atom (described by the Pauli
raising and lowering operators $\sigma_{+}$ and $\sigma_{-}$). This
interaction must then be engineered so that it takes the bilinear form%

	\begin{equation}
		\chi\left(  AS_{+}+A^{\dag}S_{-}\right)  , \label{B}%
	\end{equation}
	with $\chi$ being an effective atom-field coupling and $A^{\dag},A$
	($S_{+},S_{-}$) defined by a canonical transformation on the original
	operators for the cavity mode: $a^{\dag},a$ (auxiliary atom: $\sigma
	_{+},\sigma_{-}$ ), with the central requirement $A\left\vert \Psi
	\right\rangle =0$. The master equation for the cavity mode, coming from
	reservoir engineering, is given by%

\begin{align}
	\dot{\rho}=\left(  \Gamma/2\right)  \left(  2A\rho A^{\dag}-A^{\dag}A\rho-\rho
	A^{\dag}A\right) \nonumber\\ +\left(  \tilde{\Gamma}/2\right)  \left(  2a\rho a^{\dag
	}-a^{\dag}a\rho-\rho a^{\dag}a\right)  , \label{R}%
\end{align}

with the assumption $\Gamma\propto\chi\gg\tilde{\Gamma}$. It is therefore
clear that the Lindbladian for $a^{\dag},a$ acts as a perturbation over that
for $A^{\dag},A$, causing the fidelity of the protected state, necessarily an
eigenstate of $A$ with null eigenvalue \cite{PCZ} ($A\left\vert \Psi
\right\rangle =0$), to be slightly less than unity, since $\mathcal{F}%
\propto1-\tilde{\Gamma}/\Gamma$.

Compared to the method of state protection through engineered reservoir, the
conventional laser mechanism is far more generous since the stringent
requirement $A\left\vert \Psi\right\rangle =0$ is relaxed, with the laser
coherent steady state $\left\vert \alpha\right\rangle $ being an eigenstate of
$a$ with non null eigenvalue, $a\left\vert \alpha\right\rangle =\alpha
\left\vert \alpha\right\rangle $. Nonetheless, this relaxed condition implies
the coherence loss of the laser stationary state due to phase diffusion. Our
squeezed vacuum laser, however, despite based on the requirement $A\left\vert
\Psi\right\rangle =0$, is also more generous than the engineered reservoir
method, since the Lindbladian for $a^{\dag},a$ ---which acts in the engineered
reservoirs to prevent the fidelity from being equal to the unit--- is absent
from the laser master equation. We must return to this interesting point later.

Our strategy here is to bring the reservoir engineering method into the laser
mechanism, aiming to produce a steady squeezed vacuum state preserving phase
coherence. From the reservoir engineering we must implement the bilinear
interaction of the cavity mode with the auxiliary system, now the laser active
medium. By its turn, from the laser theory we subject the active medium to a
linear amplification process which, due to its interaction with the cavity
mode, feeds it by stimulated emission concomitantly producing a saturation
which\textbf{ }helps constructing the far-from-equilibrium steady state. Then,
by adding together the reservoir engineering technique to the laser theory, we
should be able to build a laser with zero diffusion or zero line width, and
therefore a decoherence-free state of the cavity field.

As it should be clear in the next section, the effective bilinear Hamiltonian
required for building up the squeezed vacuum laser is a particular form of
that in Eq. (\ref{B}), given by%
\begin{equation}
	\chi\left(  A\sigma_{+}+A^{\dag}\sigma_{-}\right)  , \label{b}%
\end{equation}
which suggests an isomorphism between this interaction and the Jaynes-Cummings
model, a map between the field operators $a\leftrightarrow A$ and $a^{\dag
}\leftrightarrow A^{\dag}$. This isomorphism would allow us to derive the
equations of our \textit{effective laser }directly from those of the
conventional laser theory. For the construction of such an isomorphism
relation we must derive a vector basis for the cavity field, $\left\{
\left\vert n\right\rangle _{A}\right\}  $, on which the action of our
generalized operators $\left(  A^{\dag},A\right)  $ emulate that of the usual
creation and annihilation operators $\left(  a^{\dag},a\right)  $ in the Fock
space $\left\{  \left\vert n\right\rangle \right\}  $. With this, we
immediately derive the generalized master equation that describes, in the
above-threshold regime, the construction of the steady squeezed vacuum state.
In short, all we have to do is to establish the isomorphism between the cavity
field operators in our squeezed vacuum laser and in the conventional coherent
state laser; given the isomorphism, the equations for our laser are
automatically settled.

We stress here that the constructed isomorphism between the field operators in
the effective and the Jaynes-Cummings Hamiltonians automatically results in
the engineered Lindbladian $\left(  \Gamma/2\right)  \left(  2A\rho A^{\dag
}-A^{\dag}A\rho-\rho A^{\dag}A\right)  $, differently from what happens\ in
the original reservoir engineering protocol \cite{PCZ}, in which a set of
approximations is required to obtain the desired Lindbladian. In other words,
the engineered Lindbladian comes as a gift from the constructed isomorphism,
avoiding the set of approximation imposed by the original reservoir
engineering method. Furthermore, in the derivation of the engineered
Lindbladian through the constructed isomorphism, we have the advantage of
eliminating the unwanted Lindbladian for $a^{\dag},a$ which, as observed
above, acts as a perturbation over that for $A^{\dag},A$ (causing the fidelity
of a protected state to be slightly less than unity). Although the method we
have used is not exactly that of reservoir engineering as presented in
\cite{PCZ}, all the main ingredientes of the latter method are actually
present in our conctruction: the engineering of the effective atom-field
interation and, consequently, of the associated Lindbladian.

A laser with zero linewidth is most useful for a variety of application, among
which we mention optical sensing, metrology, higher order coherent
communication, high-precision detection, and laser spectroscopy \cite{Ch}.
Therefore, the method here presented can contribute to or inspire the design
of lasers with exceedingly small diffusion and linewidth, with broad
technological applications.

Squeezed states are most efficiently generated from optical parametric
down-conversion in a non-linear $\chi^{\left(  2\right)  }$ crystal
\cite{Kimble,M}. We also mention the generation of squeezed states by four
wave mixing in an optical cavity \cite{Slusher,M}. However, our purpose here
is to demonstrate the possibility of generating squeezed state of light
through the laser mechanism, with the required nonlinearity being constructed
through the atom-field interaction itself. The squeezing of cavity-field
states through their effective interaction with atoms have been systematically
pursued in cavity quantum electrodynamics \cite{Mil}.

Our paper is organized as follows: In Section II we present a scheme, based on
the adiabatic elimination of fast variables, for the construction of the
effective Hamiltonian required for the operation of the squeezed vacuum laser.
In Section III we construct the isomorphism between the cavity field operators
in the effective and the Jaynes-Cummings interactions, and in Section IV we
present the master equation for the squeezed vacuum laser and the numerical
analysis demonstrating the effectiveness of our method. Finally, in Section V
we present our conclusions.

\section{The effective atom-field interaction}
\begin{figure}
	\centering \includegraphics[width=1\linewidth]{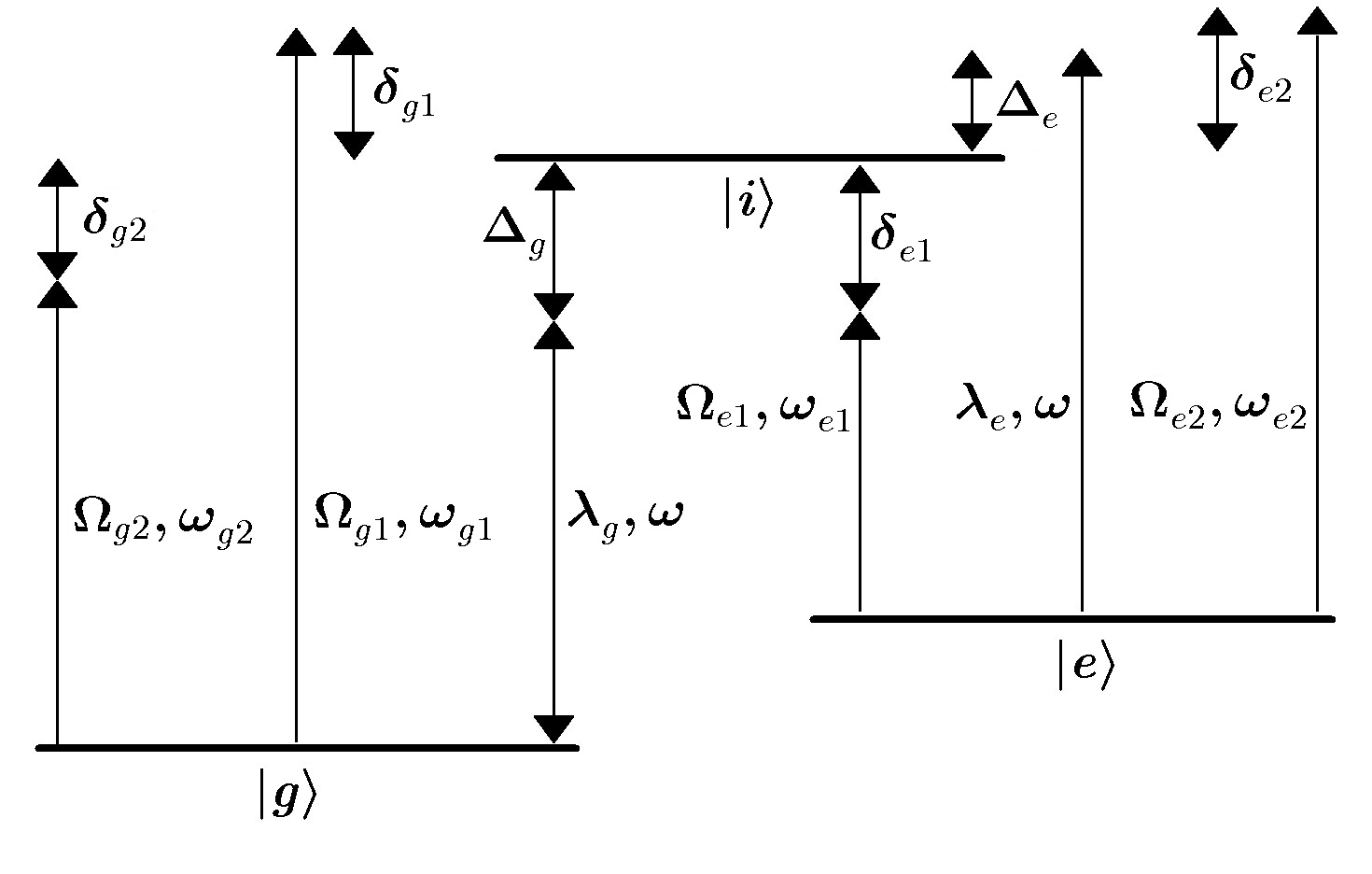}
	\caption{\textbf{Sketch of the $\mathbb{\varLambda}$ configuration to engineer the effective
			interaction for building up the squeezed vacuum laser}. The cavity mode ($\omega$) is used to promote the Raman-type transitions $g$~$\leftrightarrow$~$i$ and
			$e$~$\leftrightarrow$~ $i$, with detunings $\Delta_{g}=\omega_{i}-\omega$ and
			$\Delta_{e}=\omega_{0}+\omega-\omega_{i}$, and coupling strengths $\lambda
			_{g}$ and $\lambda_{e}$. Two pairs of laser beams ($\omega_{g\ell}$ and
			$\omega_{e\ell}$, $\ell=1,2$)  with detunings $\delta_{g1}=\omega_{g1}-\omega_{i}$, $\delta_{g2}=\omega_{i}	-\omega_{g2}$, $\delta_{e1}=\omega_{i}-\omega_{0}-\omega_{e1}$ and
			$\delta_{e2}=\omega_{0}+\omega_{e2}-\omega_{i}$ and coupling strengths
			$\Omega_{g\ell}$ and $\Omega_{e\ell}$ }
	\label{fig:fig1}
\end{figure}
The first step for achieving our\textit{ }goal is to engineer the atom-field
interaction through which we implement the amplification-saturation mechanism
building up and sustaining our squeezed vacuum state. The effective
interaction follows from considering the transitions induced by quantum and
classical fields in a three-level Lambda-type configuration, as depicted in
Fig.~\ref{fig:fig1}. The intermediate (more-excited) atomic\textbf{ }level\textbf{
}$\left\vert i\right\rangle $\textbf{ }must be considered, apart from the
lasing levels\textbf{ }$\left\vert g\right\rangle $\textbf{ }and\textbf{
}$\left\vert e\right\rangle $. The cavity mode ($\omega$) is used to promote
the Raman-type transitions $g$ $\longleftrightarrow$ $i$ and
$e\longleftrightarrow i$, with detunings $\Delta_{g}=\omega_{i}-\omega$ and
$\Delta_{e}=\omega_{0}+\omega-\omega_{i}$, and coupling strengths $\lambda
_{g}$ and $\lambda_{e}$. Two pairs of laser beams ($\omega_{g\ell}$ and
$\omega_{e\ell}$, $\ell=1,2$) help to excite the same atomic transitions with
detunings $\delta_{g1}=\omega_{g1}-\omega_{i}$, $\delta_{g2}=\omega_{i}%
-\omega_{g2}$, $\delta_{e1}=\omega_{i}-\omega_{0}-\omega_{e1}$ and
$\delta_{e2}=\omega_{0}+\omega_{e2}-\omega_{i}$ and coupling strengths
$\Omega_{g\ell}$ and $\Omega_{e\ell}$. The Hamiltonian describing the process
is given by $H(t)=H_{0}+V(t)$, where
\begin{subequations}
\label{1}%
\begin{align}
 H_{0}=& \omega a^{\dag}a+\omega_{0}\sigma_{ee}+\omega_{i}\sigma
_{ii}\text{,}\label{1a}\\
 V(t)=&\lambda_{g}a\sigma_{ig}+\lambda_{e}a\sigma_{ie}+\nonumber\\
&{\textstyle\sum\nolimits_{\ell}}
\left(  e^{-i\omega_{g\ell}t}\Omega_{g\ell}\sigma_{ig}+e^{-i\omega_{e\ell}%
t}\Omega_{e\ell}\sigma_{ie}\right)  +h.c.\text{,} \label{1b}%
\end{align}
using the Pauli operators $\sigma_{rs}=\left\vert r\right\rangle \left\langle
s\right\vert $, with $r,s$ denoting the atomic levels. In what follows we
assume $\lambda_{g}$ $=$ $\lambda_{e}=\lambda$, $\Omega_{g1}=$ $\Omega_{g2}=-$
$\Omega_{e1}=-$ $\Omega_{e2}$, $\delta_{g1}=\delta_{g2}=\delta_{e1}%
/\kappa=\delta_{e2}/\kappa$, $\Delta_{g}=\delta_{e1}$ and $\Delta_{e}%
=\delta_{g1}$. Then, under the set of parameter $\delta_{r\ell}\gg
\Omega_{r\ell}\gg\bar{n}\lambda$, with $\bar{n}$ being the average photon
number in the cavity, we verify that, in the interaction picture, the
non-diagonal Hamiltonian $\mathcal{H}(t)$ consists of highly oscillatory terms
such that, to a good approximation, we obtain the second-order effective
Hamiltonian \cite{Methods1}%
\end{subequations}
\begin{equation}
H_{eff}=-i\mathcal{H}(t)\int_{0}^{t}d\tau\mathcal{H}(\tau)=g\left(
A\sigma_{+}+A^{\dag}\sigma_{-}\right)  \text{,} \label{3}%
\end{equation}
where $\sigma_{+}=|e\rangle\langle g|$, $\sigma_{-}=|g\rangle\langle e|$ and
the coupling strength is given by $g=\sqrt{1-\kappa^{2}}\lambda\Omega
_{g1}/\delta_{e1}$, with $\kappa=\delta_{e1}/\delta_{g1}$, and the generalized
operators read
\begin{equation}
A=\frac{a+\kappa a^{\dag}}{\sqrt{1-\kappa^{2}}}\text{, \ \ }A^{\dag}%
=\frac{a^{\dag}+\kappa a}{\sqrt{1-\kappa^{2}}}\text{,} \label{4}%
\end{equation}
with $\left[  A,A^{\dag}\right]  =\left[  a,a^{\dag}\right]  =1$.

In order to verify the validity of the approximations leading from $V(t)$ to
$H_{eff}$, we plot in Figs. 2(a) and (b) the variances of the quadratures
$X_{1}=\left(  a^{\dagger}+a\right)  /2$ and $X_{2}=\left(  a-a^{\dagger
}\right)  /2i$ for the field states generated by both the full and the
effective Hamiltonians in Eqs. (\ref{1}) and (\ref{3}), against $gt$,
i.e. the number of cycles of the effective coupling. We start with the atom
in the excited state $\left\vert e\right\rangle $ and the cavity mode in the
vacuum state $\left\vert 0\right\rangle $, considering, in units of the Rabi
frequency $\lambda$, the parameters $\delta_{g1}=10^{3}$, $\delta_{e1}%
=6\times10^{2}$, $\Omega_{g1}=40$, and $g=0.05$, such that $\kappa=0.6$. In
Figs. 2(a) and (b) the straight and dotted lines refer to the effective and
full Hamiltonians, respectively, showing a good agreement between both curves
up to $gt\approx7$. We have also plotted in Figs. 2(c) and (d) the
excitations $\left\langle a^{\dagger}a\right\rangle (t)$ and $\left\langle
\sigma_{ee}\right\rangle (t)$ against $gt$, and we again see a very good
agreement between the curves generated by the full and the effective
Hamiltonians until the same $gt\approx7$.

Regarding the engineering of the effective atom-field interaction (\ref{3}), a
detailed account on Raman transition in cavity quantum electrodynamics can be
found in Ref. \cite{Boozer}. We note that the atomic level configuration we
have used to engineer the required interaction is certainly not unique; it can
be engineered from other level configuration using more or less classical
fields. Finally, we stress that in engineering the effective Hamiltonian
(\ref{3}) we have not take into account the usually small amplitude and phase
fluctuations of the required laser beams, which would indeed result in some
phase diffusion of our squeezed vacuum laser.
At this point we mention that another proposal to achieve squeezed lasing has
been presented in which the cavity is parametric driven with the help of a
non-linear $\chi^{\left(  2\right)  }$ crystal inside the cavity \cite{MJ}.
Our engineered atom-field interaction thus repalces the parametric driven
process in Ref. \cite{MJ}, dispensing the non-linear crystal inside the
cavity\ and the coherent drive of the cavity mode. However, since our laser
requires the effective atom-field interaction, it present an operating
timescale after which it must be restarted.

\section{The isomorphism between the $A^{\dag},A$ and $a^{\dagger},a$
algebras}

Having engineered the required interaction (\ref{3}), we now start to
construct the vector basis $\left\{  \left\vert n\right\rangle _{A}\right\}  $
for the cavity field, in whose states $\left\vert n\right\rangle _{A}$ the
action of operators $A^{\dag}A$, $A^{\dag}$ and $A$, must lead to the same
relations as those resulting from the actions of $a^{\dag}a,$ $a^{\dag}$ and
$a$ on the Fock basis $\left\{  \left\vert n\right\rangle \right\}  $, i.e.:
\begin{subequations}
\label{10}%
\begin{align}
A^{\dag}A\left\vert n\right\rangle _{A}  &  =n\left\vert n\right\rangle
_{A},\label{10a}\\
A^{\dag}\left\vert n\right\rangle _{A}  &  =\sqrt{n+1}\left\vert
n+1\right\rangle _{A},\label{10b}\\
A\left\vert n\right\rangle _{A}  &  =\sqrt{n}\left\vert n-1\right\rangle _{A}.
\label{10c}%
\end{align}
All the basis states $\left\{  \left\vert n\right\rangle _{A}\right\}  $ are
constructed from the vacuum state $\left\vert 0\right\rangle _{A}$, starting
from the relation
\end{subequations}
\begin{equation}
A\left\vert 0\right\rangle _{A}=0\text{,} \label{11}%
\end{equation}
which enables us to determine the probability amplitudes $c_{n}$ defining the
superposition $\left\vert 0\right\rangle _{A}=\sum\nolimits_{n}c_{n}\left\vert
n\right\rangle $. Considering the operator $A$ as given by Eq. (\ref{4}), we
first compute the vacuum state from Eq. (\ref{11}) and then, using the
relation $\left(  A^{\dag}\right)  ^{n}\left\vert 0\right\rangle _{A}%
/\sqrt{n!}$, we derive all the even and odd generalized excitations, given by
\begin{subequations}
\label{12}%
\begin{align}
 & \left\vert 2m\right\rangle _{A}=\frac{\left(  1-\kappa^{2}\right)
^{1/4}}{\sqrt{\left(  2m\right)  !}}\sum\limits_{n=0}^{\infty}\left(
-\kappa\right)  ^{n-m}\sqrt{\frac{\left(  2n-1\right)  !!}{\left(  2n\right)
!!}}\nonumber\\
&  \times\sum\limits_{\ell=0}^{m}\binom{m}{\ell}\left(  -\kappa^{2}\right)
^{\ell}\frac{\left(  2n\right)  !!}{\left[  2\left(  n+\ell-m\right)  \right]
!!}\frac{\left[  2\left(  n+\ell\right)  -1\right]  !!}{\left(  2n-1\right)
!!}\left\vert 2n\right\rangle \text{,}\label{12a}\\
&  \left\vert 2m+1\right\rangle _{A}=\frac{\left(  1-\kappa^{2}\right)
^{3/4}}{\sqrt{\left(  2m+1\right)  !}}\sum\limits_{n=0}^{\infty}\left(
-\kappa\right)  ^{n-m}\sqrt{\frac{\left(  2n+1\right)  !!}{\left(  2n\right)
!!}}\nonumber\\
&  \times\sum\limits_{\ell=0}^{m}\binom{m}{\ell}\left(  -\kappa^{2}\right)
^{\ell}\frac{\left(  2n\right)  !!}{\left[  2\left(  n+\ell-m\right)  \right]
!!}\frac{\left[  2\left(  n+\ell\right)  +1\right]  !!}{\left(  2n+1\right)
!!}\left\vert 2n+1\right\rangle \text{.} \label{12b}%
\end{align}
\end{subequations}

The basis defined by the even and odd number states given by Eqs. (\ref{12}),
together with the laser model establishes the isomorphism
between the  laser field generated by the interaction from Eq. 3,
written in the new basis, and the usual laser field, in the Fock basis, since
their master equations are similar, containing each the Liouville-Von Neumman
term and the cavity dissipative term in the Lindblad form, referring to the
loss of an Harmonic Oscillator (HO) to the enviroment. The Liouville-Von
Neumman term would be equal in both lasers, once we use the Hamiltonian in the
new basis, as in Eq. 3, and since now we have a system guided by an
interaction in the form of Jaynes-Cummings, as describing an HO, naturally we
can say that the cavity dissipates in the Lindblad form. Knowing that the steady state of the conventional laser is
the coherent state $\left\vert \alpha\right\rangle $ (owing to the
Jaynes-Cummings atom-field interaction), it is then automatic to derive the
steady state of our laser, which results from the effective Hamiltonian
(\ref{3}). Once the isomorphism is established, all we have to do is to
describe the coherent state in the vector basis $\left\{  \left\vert
n\right\rangle _{A}\right\}  $, i.e., $\left\vert \alpha\right\rangle
_{A}=D_{A}(\alpha)\left\vert 0\right\rangle _{A}=\exp\left(  \alpha A^{\dag
}-\alpha^{\ast}A\right)  \left\vert 0\right\rangle _{A}=e^{-\left\vert
\beta\right\vert ^{2}/2}e^{\beta a^{\dag}}e^{-\beta^{\ast}a}\left\vert
0\right\rangle _{A}$, with $\beta=\left(  \alpha-\kappa\alpha^{\ast}\right)
/\sqrt{1-\kappa^{2}}$. We obtain, expanded in the usual Fock basis $\left\{
\left\vert n\right\rangle \right\}  $, the state%
\begin{eqnarray}
	\left\vert \alpha\right\rangle _{A}=(1-\kappa^{2})^{1/4}e^{\left(
		\kappa\alpha^{2}-\left\vert \alpha\right\vert ^{2}\right)  /2}\cr\sum
	\limits_{n=0}^{\infty}\sqrt{n!}\left (  \sum\limits_{\ell=0}^{n}\left(
	\frac{\kappa}{2}\right)  ^{\ell/2}\frac{\beta^{n-\ell}}{\left(  n-\ell\right)
		!\ell!}H_{\ell}\left(  x\right)  \right)  \left\vert n\right\rangle
	,\label{13}%
\end{eqnarray}
\begin{figure*}
	\centering
	\begin{subfigure}{0.45\textwidth}
		\centering
		\includegraphics[width=\textwidth]{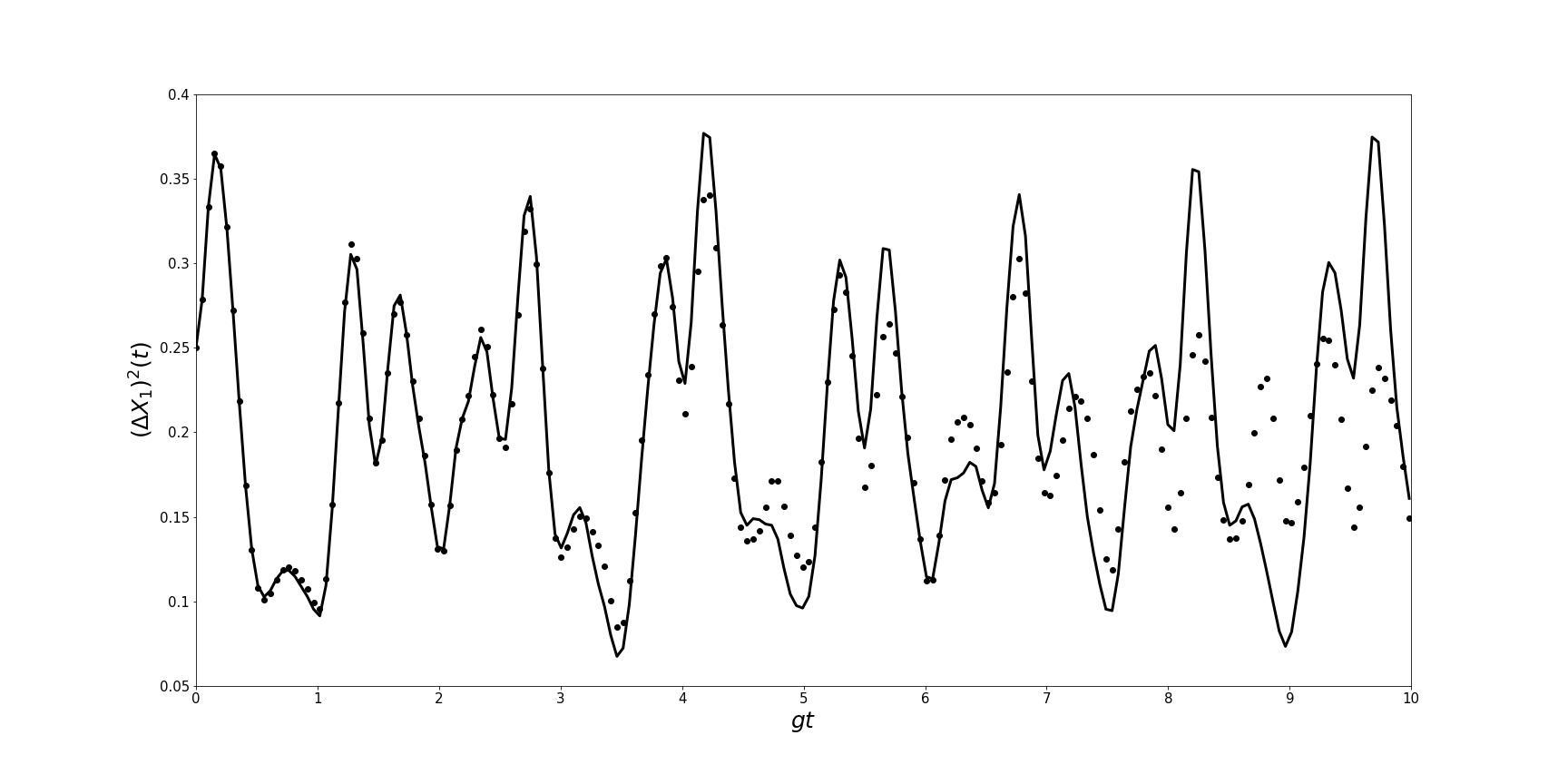}
		\caption{}
		\label{fig:a}
	\end{subfigure}
	\hfill
	\begin{subfigure}{0.45\textwidth}
		\centering
		\includegraphics[width=\textwidth]{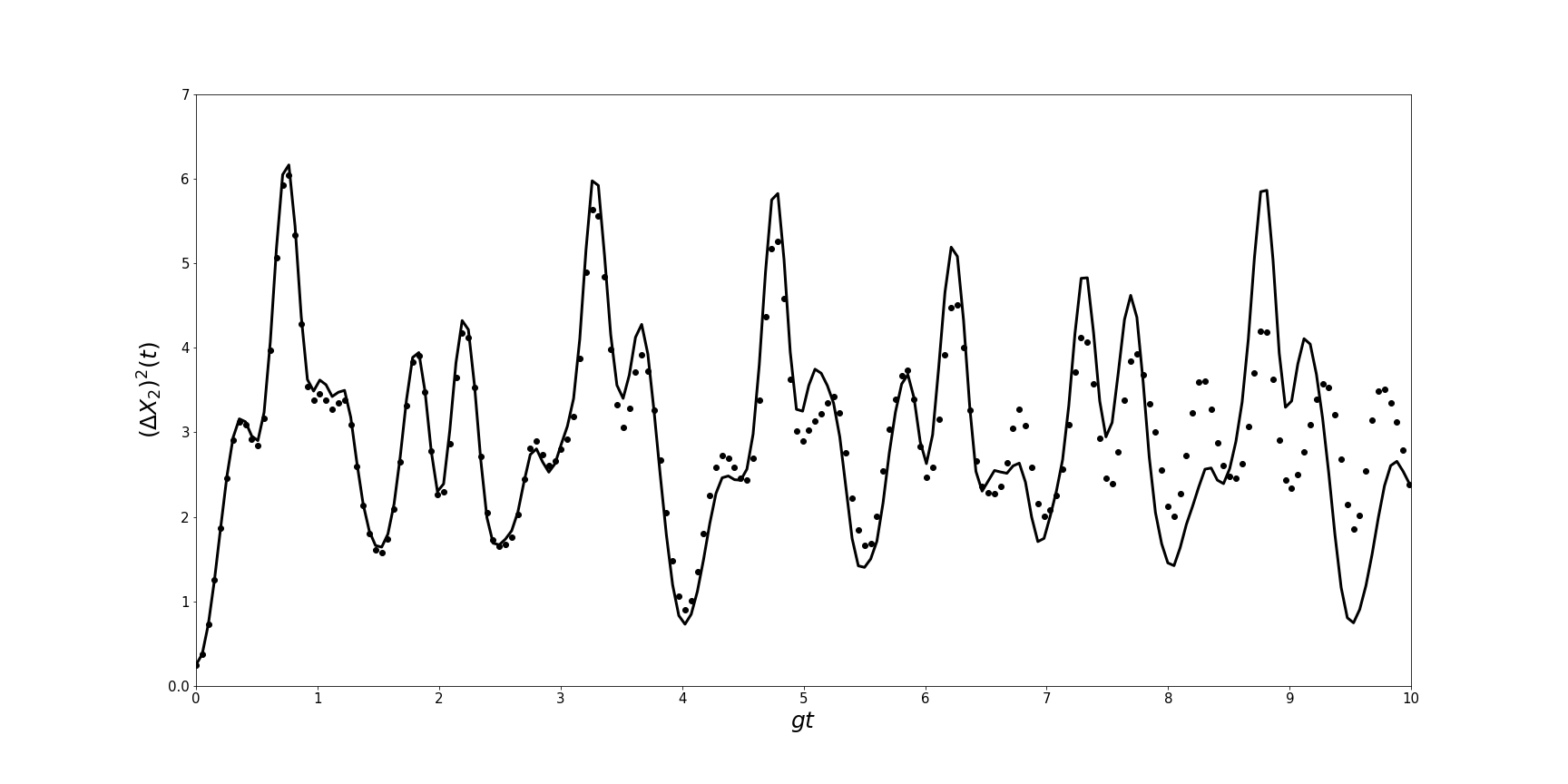}
		\caption{}
		\label{fig:b}
	\end{subfigure}
	\hfill
	\begin{subfigure}{0.45\textwidth}
		\centering
		\includegraphics[width=\textwidth]{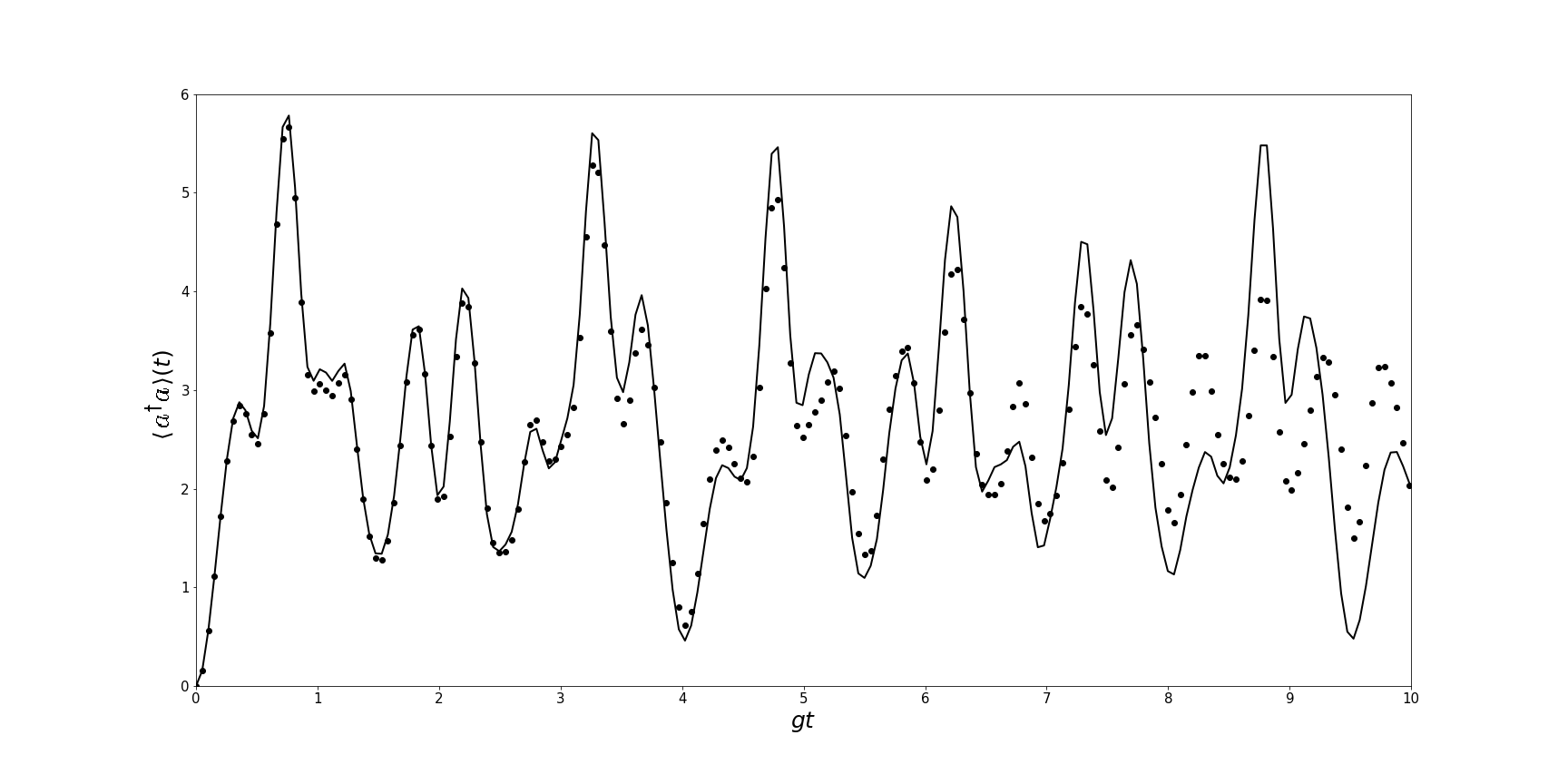}
		\caption{}
		\label{fig:c}
	\end{subfigure}
	\hfill
	\begin{subfigure}{0.45\textwidth}
		\centering
		\includegraphics[width=\textwidth]{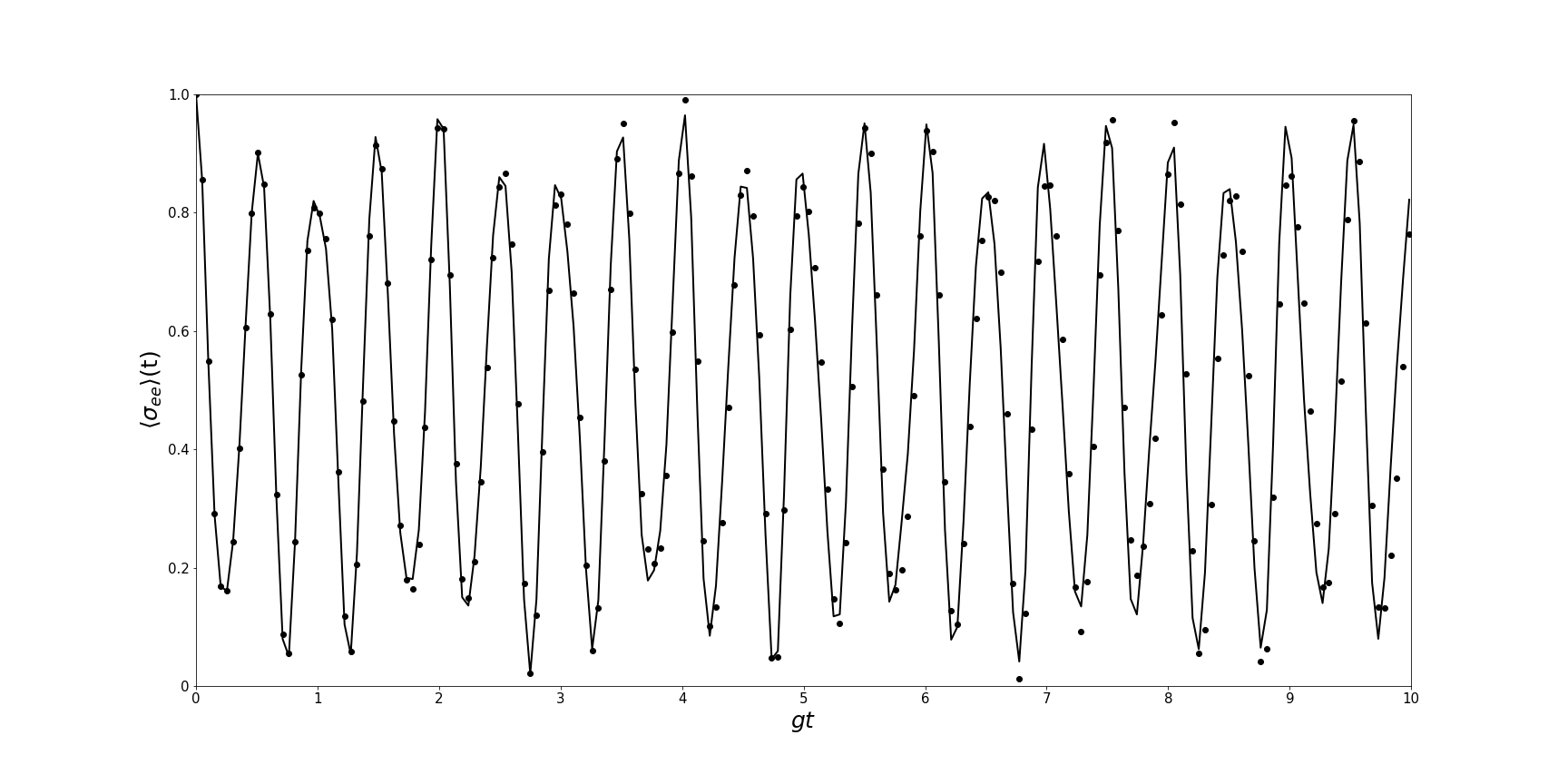}
		\caption{}
		\label{fig:d}
	\end{subfigure}
	\caption{\textbf{Validity of the Effective interaction}. The the variances \textbf{(a)}, $\left(  \Delta X_{1}\right)
		^{2}(t)$ and \textbf{(b)}, $\left(  \Delta X_{2}\right)  ^{2}(t)$, and the excitations
		\textbf{(c)}, $\left\langle a^{\dagger}a\right\rangle (t)$ and \textbf{(d)}, $\left\langle
		\sigma_{ee}\right\rangle (t)$ in function of $gt$, for the field states generated by the effective
		and full Hamiltonians (straight and dotted lines, respectively). We have
		started with the atom and the cavity mode in the excited and the vacuum state,
		$\left\vert e\right\rangle \left\vert 0\right\rangle $, considering, in units
		of the Rabi frequency $\lambda$, the parameters $\delta_{g1}=10^{3}$,
		$\delta_{e1}=6\times10^{2}$, $\Omega_{g1}=40$, and $g=0.05$, such that
		$\kappa=0.6$.}
	\label{fig:fig2}
\end{figure*}

\begin{figure}
	\centering
	\begin{subfigure}{0.45\textwidth}
		\centering
		\includegraphics[width=\textwidth]{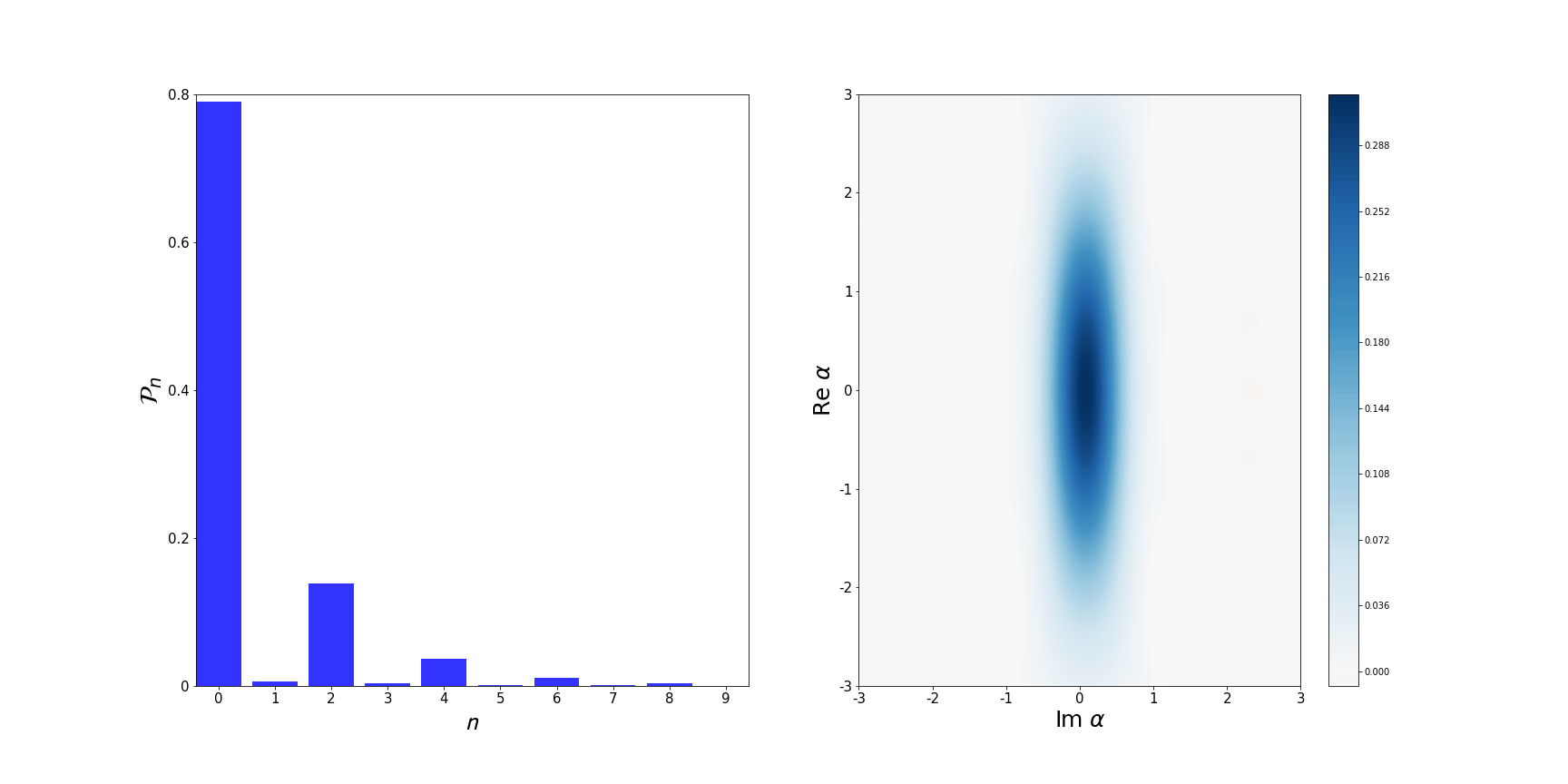}
		\caption{}
		\label{fig:a}
	\end{subfigure}
	\hfill
	\begin{subfigure}{0.45\textwidth}
		\centering
		\includegraphics[width=\textwidth]{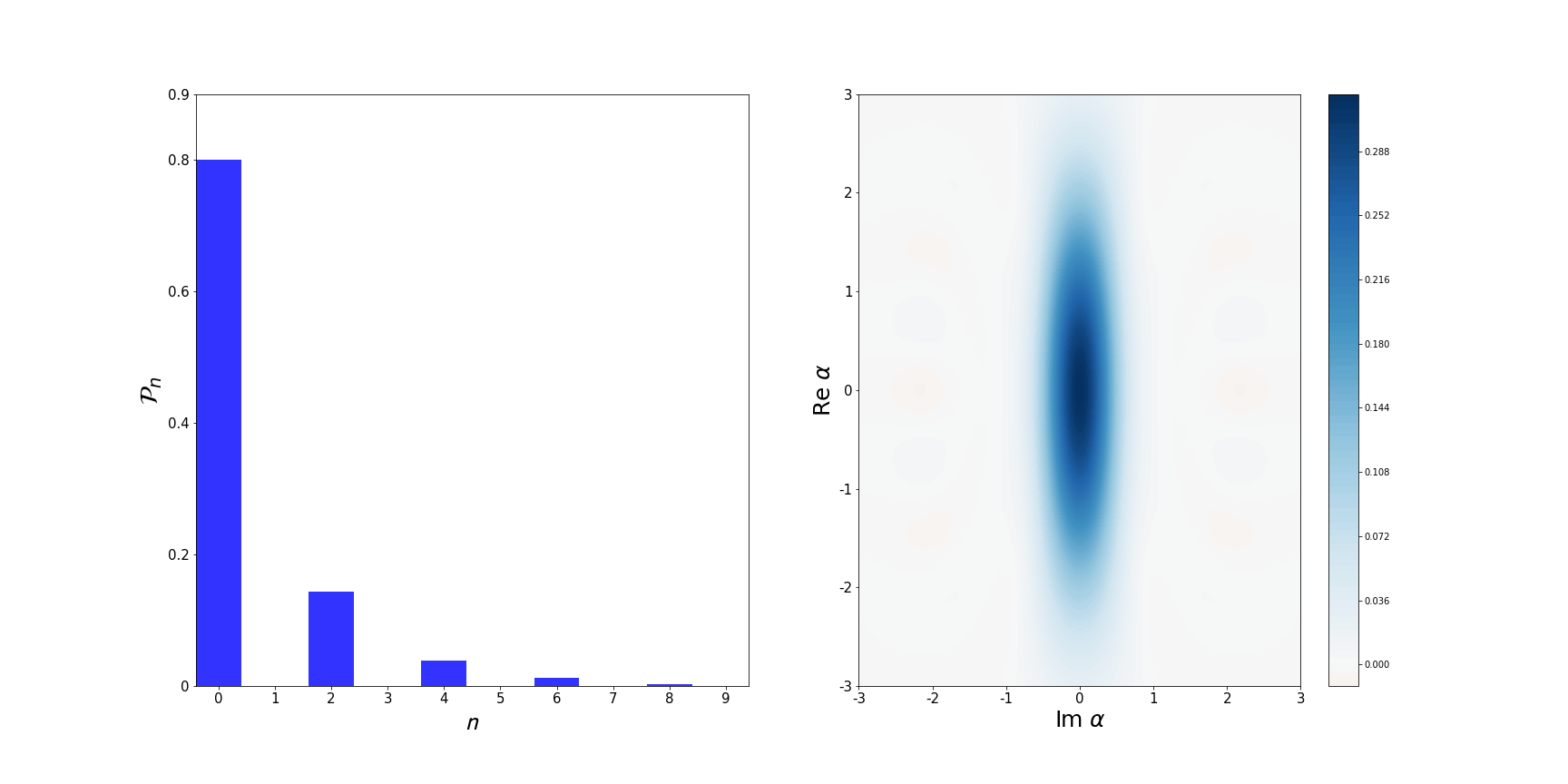}
		\caption{}
		\label{fig:b}
	\end{subfigure}
	
	\caption{\textbf{The photon number distribution $\mathcal{P}_{n}$ and the Wigner function.}\textbf{(a)}, generated by the effective Hamiltonian for $\alpha=0.18$ and $r=0.69$.\textbf{(b)}, ideal squeezed vacuum $S(r=0.69)\left\vert 0\right\rangle $.}
	\label{fig:fig3}
\end{figure}
\begin{figure}
	\centering \includegraphics[width=1\linewidth]{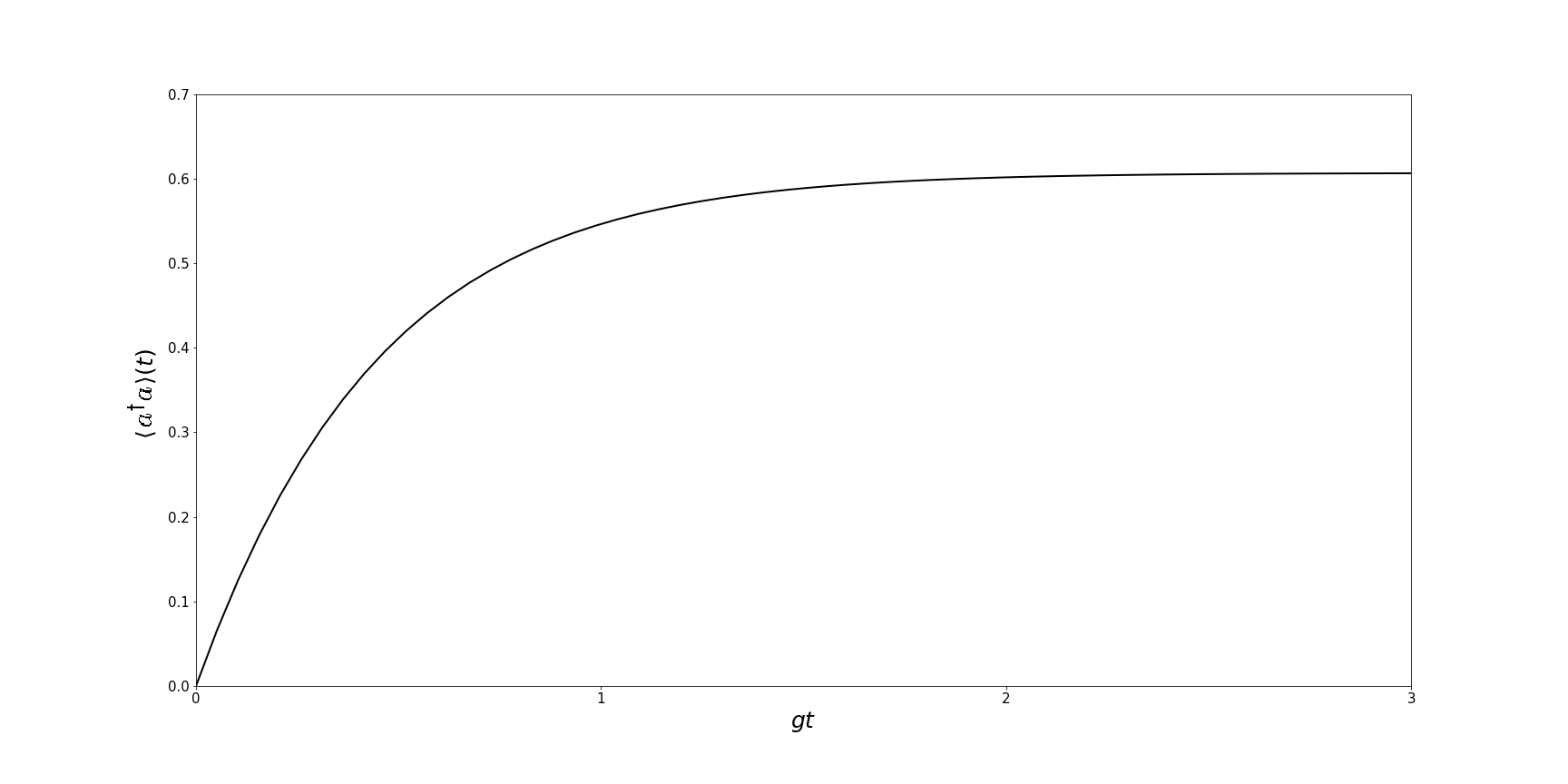}
	\caption{\textbf{The photon number}. The mean occupation number $\left\langle a^{\dag}a\right\rangle (t)$,
		considering the same parameters used in Fig. $2$, together with $\mathcal{C}%
		/g=0.35$, $r/g=92$ and $\gamma/g=0.5$, leading to $\mathcal{A}/g=736$. We
		consider the cavity initially in the vacuum state and the atoms prepared in
		their excited states. }
	\label{fig:fig4}
\end{figure}
\begin{figure}
	\centering
	\begin{subfigure}{0.45\textwidth}
		\centering
		\includegraphics[width=\textwidth]{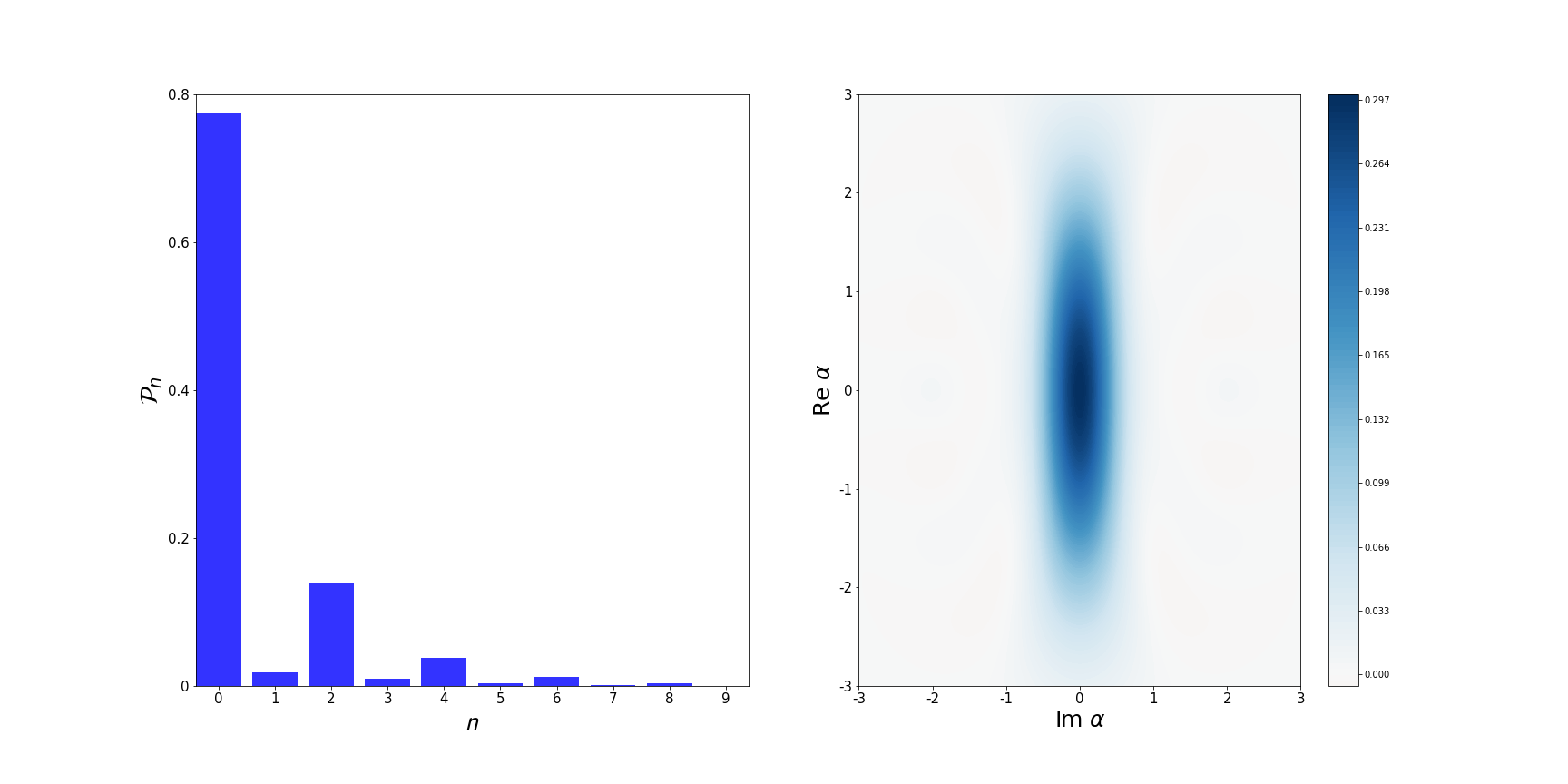}
		\caption{}
		\label{fig:a}
	\end{subfigure}
	\hfill
	\begin{subfigure}{0.45\textwidth}
		\centering
		\includegraphics[width=\textwidth]{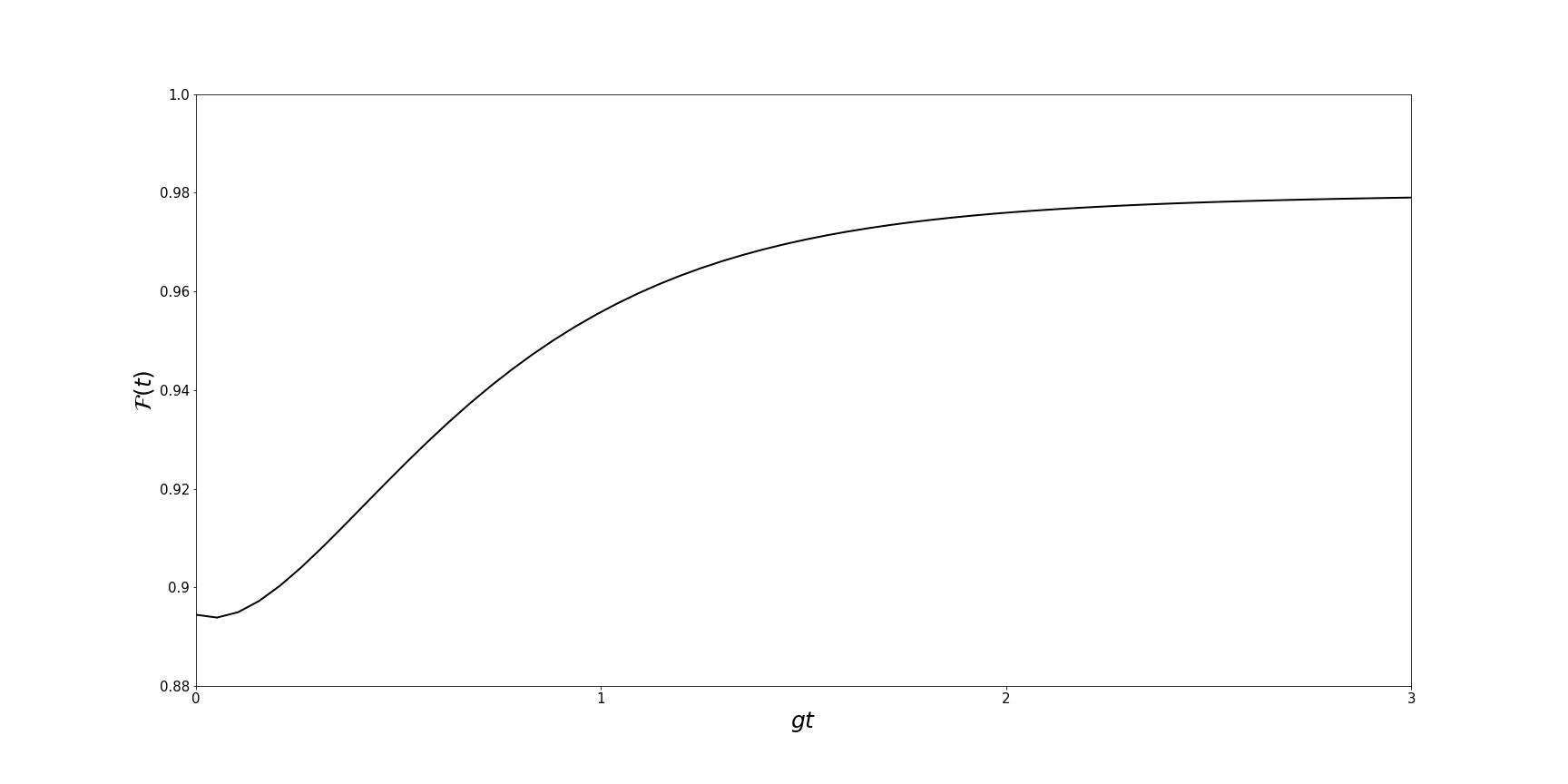}
		\caption{}
		\label{fig:b}
	\end{subfigure}
	
	\caption{\textbf{Squeezed vacuum laser}. \textbf{(a)}, The photon number distribution $\mathcal{P}_{n}$ and the phase space
		projection of the Wigner function of the produced laser state for $gt=4$, and
		\textbf{(b)} the fidelity of the evolved laser state against $gt$. We use the same
		parameters of Fig. 4.}
	\label{fig:fig5}
\end{figure}
\begin{figure}
	\centering \includegraphics[width=1\linewidth]{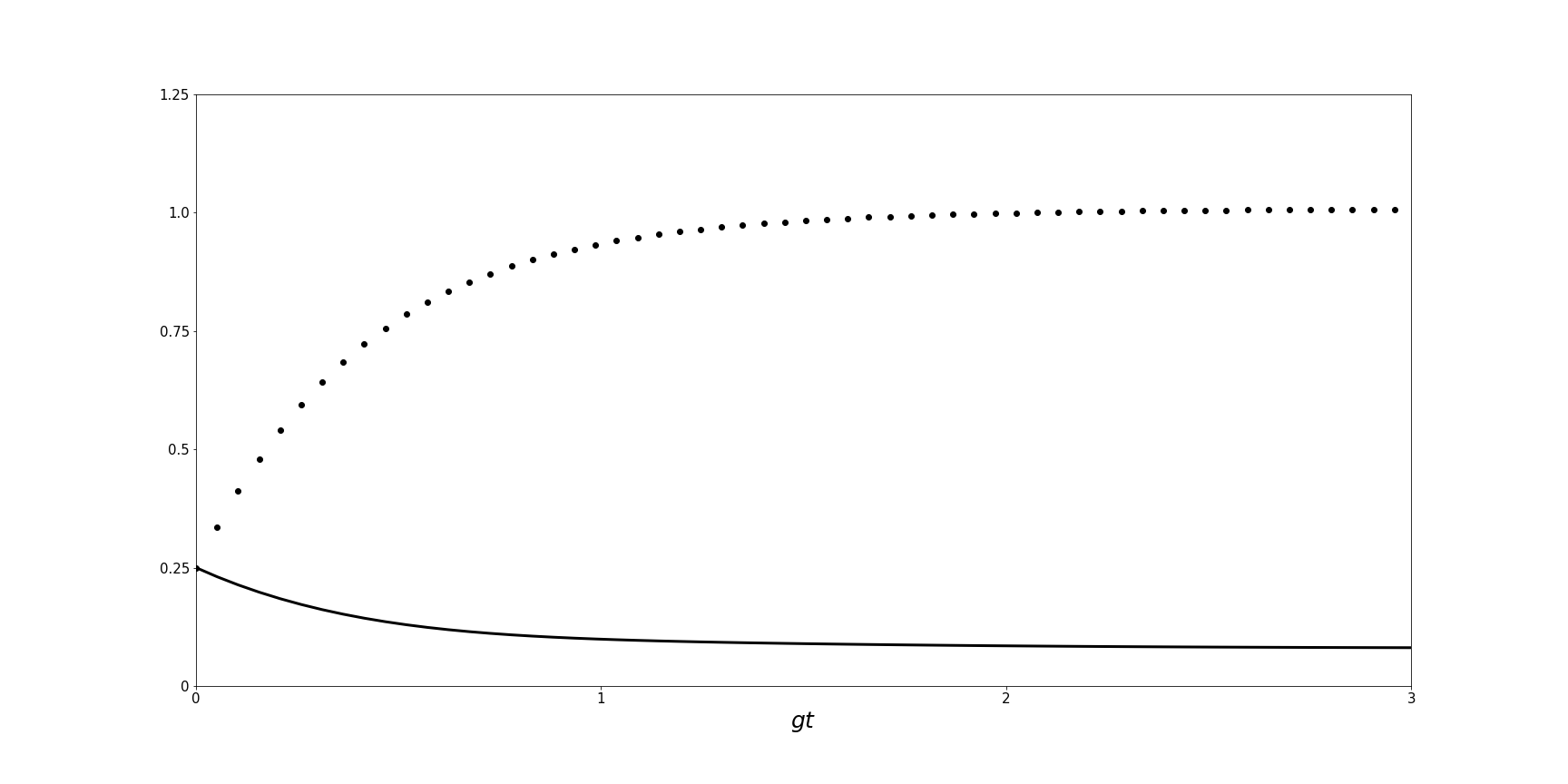}
	\caption{\textbf{The variances},  $\left(  \Delta X_{1}\right)  ^{2}(t)$ (solid line) and $\left(  \Delta X_{2}\right)  ^{2}(t)$ (dotted line) of the generated laser
		state against $gt$, using the same parameters of Fig. 4. }
	\label{fig:fig6}
\end{figure}
\begin{figure}
	\centering \includegraphics[width=1\linewidth]{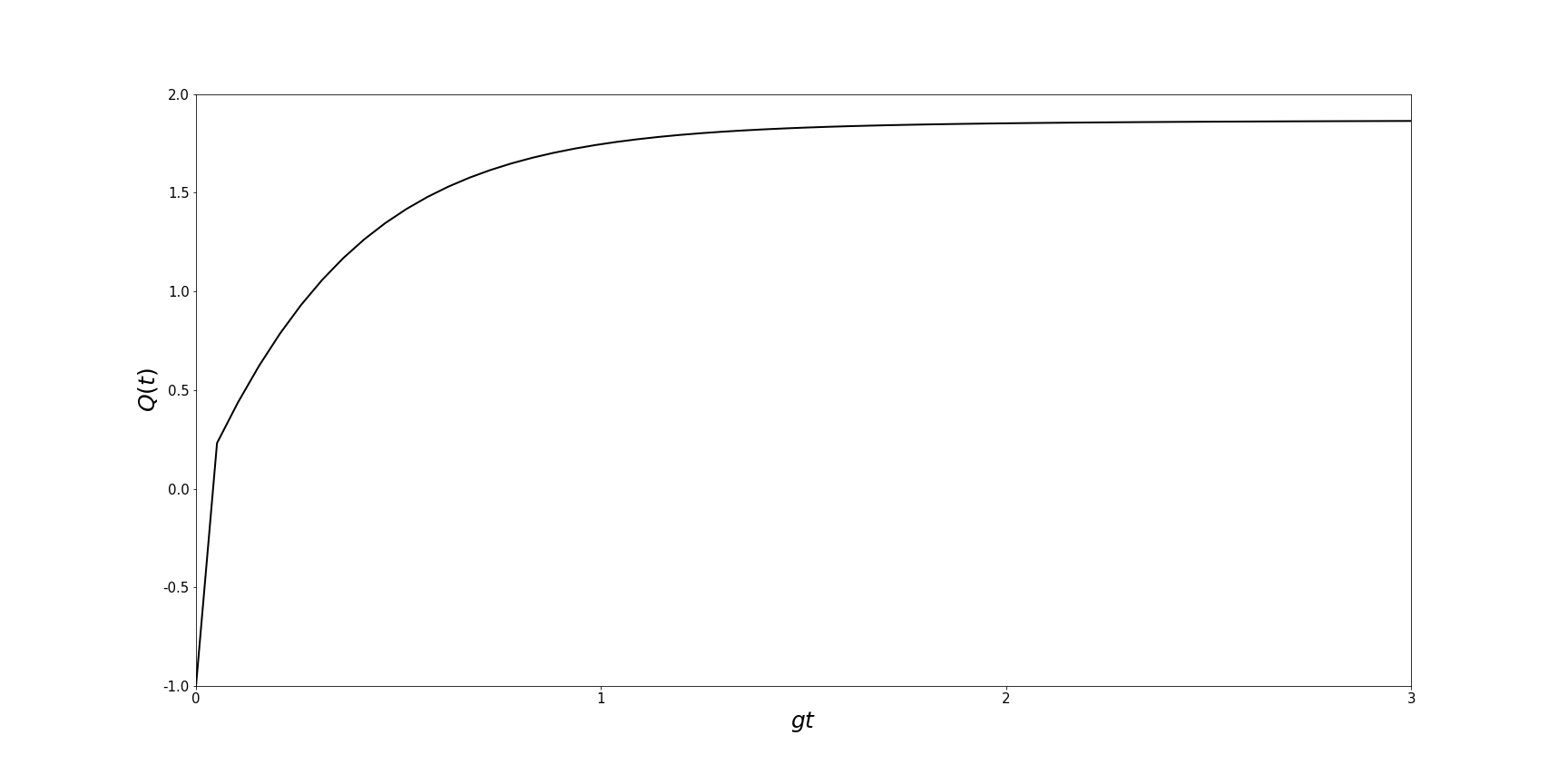}
	\caption{\textbf{The nonclassicality}.  The Mandel $Q_{M}$ parameter for the laser state against $gt$, using the same parameters of Fig. 4 }
	\label{fig:fig7}
\end{figure}
\begin{figure}
	\centering \includegraphics[width=1\linewidth]{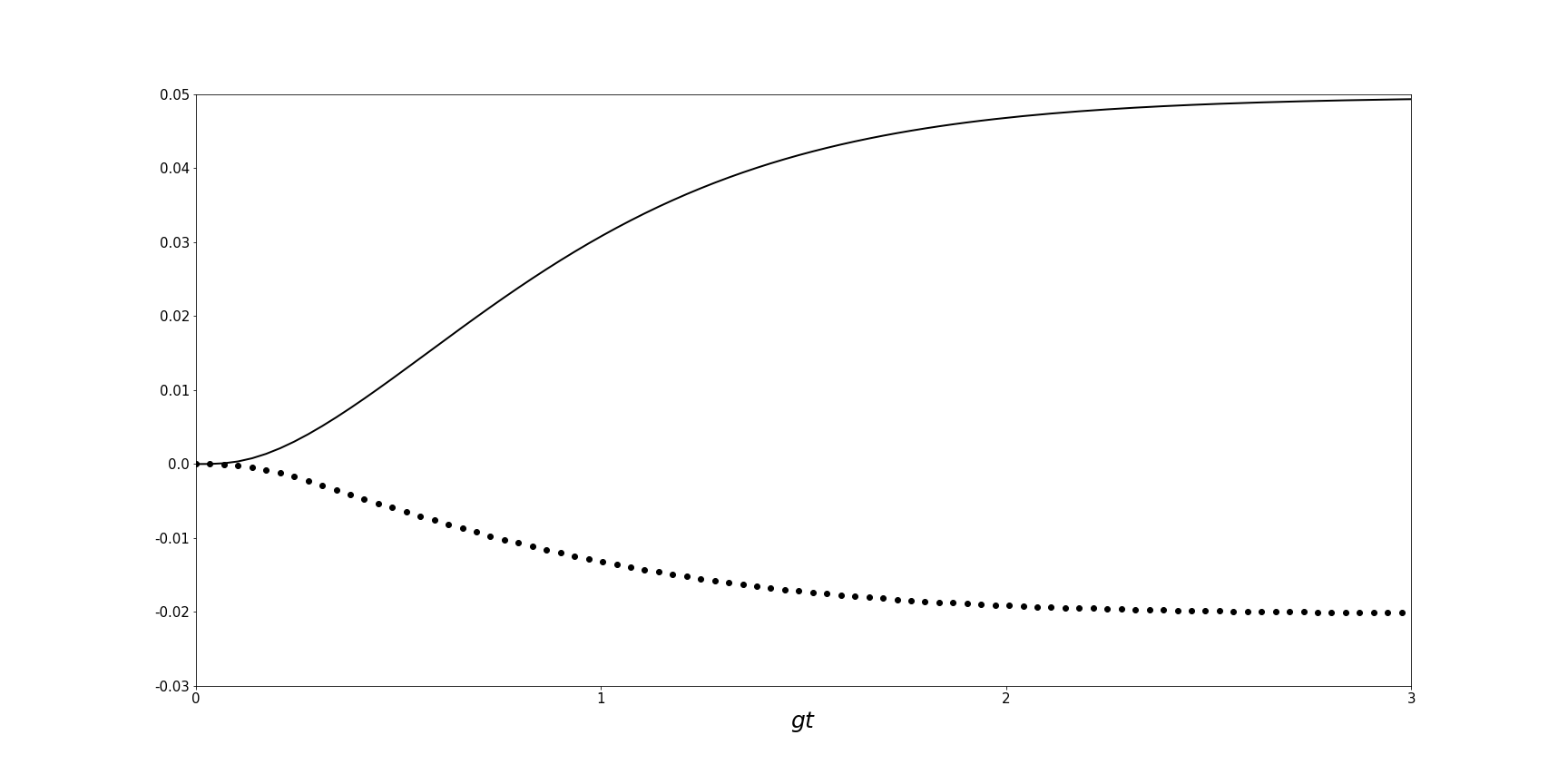}
	\caption{\textbf{The absence of diffusion}. Off-diagonal matrix elements $\rho_{0,8}(t)$ and $\rho_{4,6}(t)$ against $gt$, using the same parameters of Fig. 4. }
	\label{fig:fig8}
\end{figure}
We note that for $\kappa=0$ we immediately recover the usual Fock basis states
from Eq. (\ref{12}) and the usual coherent state from Eq. (\ref{13}). In Fig.
3(a) we show the photon number distribution $\mathcal{P}_{n}$ and a
phase-space plot of the uncertainties of the laser state given by Eq.
(\ref{13}), using\textbf{ }$\alpha=0.18$ and $r=\operatorname*{Tanh}%
^{-1}\kappa=0.69$, and considering $\kappa=0.6$ as in Fig. 2. The obtained
squeezed vacuum displays a good agreement with the ideal squeezed vacuum
$S(r=0.69)\left\vert 0\right\rangle $, whose photon distribution and
projection of the Wigner distribution in phase space is shown in Fig. 3(b),
where $S$ stands for the well-known squeeze operator
\begin{equation}
S(\xi)=\exp\left[  \xi^{\ast}a^{2}-\xi\left(  a^{\dag}\right)  ^{2}\right]
,\label{Sq}%
\end{equation}
with $\xi=re^{i\varphi}$, $r$ being the degree of squeezing and $\varphi$ the
squeezing direction in phase space \cite{M,W}, here with $\varphi=0$. The
produced laser state deviates slightly from the ideal squeezed vacuum as
indicated by the populations of the odd Fock states. We point out
that, the squeezed vacuum with squeezing factor $r=\tanh^{-1}\kappa$,
is an eigenstate of $A$ with null eigenvalue, as required by the
engineering reservoir technique.

\section{The master equation for the squeezed vacuum laser}

From the isomorphism we have established, and following the footsteps of the
conventional laser theory, we can derive the master equation describing the dynamics of the
cavity field when interacting with a pumped atomic sample, through the
effective Hamiltonian (\ref{3}), and the environment. This master equation is
given by%

\begin{equation}
\dot{\rho}=\mathcal{L}_{\mathcal{A}}\rho+\mathcal{L}_{\mathcal{B}}%
\rho+\mathcal{L}_{\mathcal{C}}\rho,\label{15}%
\end{equation}
where the Lindbladians accounting for gain, saturation and cavity loss, obey
the expressions
\begin{subequations}
\label{16}%
\begin{align}
\mathcal{L}_{\mathcal{A}}\rho &  =\frac{\mathcal{A}}{2}\left(  2A^{\dag}\rho
A-AA^{\dag}\rho-\rho AA^{\dag}\right)  ,\label{16a}\\
\mathcal{L}_{\mathcal{B}}\rho &  =\frac{\mathcal{B}}{2}\left[  \frac{1}%
{4}AA^{\dag}\left(  AA^{\dag}\rho+3\rho AA^{\dag}\right) \right.  \nonumber\\
&  \left. +\frac{1}{4}\left(
\rho AA^{\dag}+3AA^{\dag}\rho\right)  AA^{\dag}\right.  \nonumber\\
&  \left.  -A^{\dag}\left(  AA^{\dag}\rho+\rho AA^{\dag}\right)  A\right]
+...,\label{16b}\\
\mathcal{L}_{\mathcal{C}}\rho &  =\frac{\mathcal{C}}{2}\left(  2A\rho A^{\dag
}-A^{\dag}A\rho-\rho A^{\dag}A\right)  .\label{16c}%
\end{align}
\end{subequations}
The coefficients for gain $\mathcal{A}=2R\left(  g/\gamma\right)  ^{2}$,
saturation $\mathcal{B}=4\mathcal{A}\left(  g/\gamma\right)  ^{2}$, and loss
$\mathcal{C}=\omega/Q$, are defined from the atomic pumping rate $R=Kp$, $K$
being the total atomic injection rate and $p$ the probability for the atomic
laser excitation, the Rabi frequency $g$, and effective atomic decay rate
$\gamma$, and the cavity quality factor $Q$. The Lindbladian for saturation in
Eq. (\ref{16b}) is given only up to 4-th order in $g$. The isomorphism then
assures us that the squeezed vacuum state in Eq. (\ref{13}) follows directly
from the competition between amplification ($\mathcal{A}$) and dissipation
($\mathcal{C}$), mediated by the the saturation of the cavity field excitation
($\mathcal{B}$), described by the master equation (\ref{15}). Far above
threshold, when $\mathcal{A\gg C}$, the far from equilibrium steady state of
the cavity field is given by Eq. (\ref{13}).

We note here that unlike what happens with engineered reservoirs, we do not
have a term in Eq. (\ref{15}) (similar to the Lindbladian for $a^{\dag},a$ in
Eq. (\ref{R})), that acts to introduce error in the laser mechanism. This is
indeed a remarkable bonus for our method, in which the only source of errors
stems from the engineering of the effective Hamiltonian (\ref{3}). By limiting
the time interval of the laser operation such that $gt=7$, the errors coming
from the engineered protocol must, however, be small as we have seen from Fig. 2.

To further support our results coming from the isomorphism ---that the laser
resulting from the atom-field interaction described by the effective
Hamiltonian (\ref{3}) is indeed the squeezed vacuum---, in what follows we
numerically analyze the construction, step by step, of such a cavity field
steady state. To do that, we numerically simulate the passage of a
dense flux of atoms through an excitation region where each atom has a
probability $p$ of being excited to the level $|e\rangle$
before entering the cavity. Considering that this flux has a regular pump
rate, $K$, we can say that the number of atoms passing through a time $\Delta t$ is equal to $K\Delta t$, as much as the average
number of excited atoms that reach the cavity is $N=k\Delta t$, where
$k=pK$. The atoms arrive successively in the cavity so that the
atom-field interaction described by (\ref{3}) still apllies
collectively, since it has being developed individually. The first atom finds
the cavity field in its initial vacuum state $\rho(0)$, leading it to the
state $\rho(t)$, with $t=1/k$. After its passage through the cavity, we
compute the reduced density operator for the field state by tracing over the
atomic degrees of freedom. The second atom then finds the cavity in this
reduced state, leading it to another reduced state at time $2t$, and so on
until the time $Nt$, each step being described by the equation%

\begin{eqnarray}
	\dot{\rho}_{i+1}(t)=k\left[  \rho(t_{i+1})-\rho(t_{i})\right]  =-i\left[
	H_{eff}(t_{i}),\rho(t_{i})\right] \cr
	 +\left(  C/2\right)  \left[  2A\rho(t_{i})A^{\dag}-A^{\dag}A\rho
	(t_{i})-\rho(t_{i})A^{\dag}A\right] \cr
	  +\left(  \gamma/2\right)  \left(  2\sigma_{-}\rho(t_{i})\sigma_{+}%
	-\sigma_{+}\sigma_{-}\rho(t_{i})-\rho(t_{i})\sigma_{+}\sigma_{-}\right)\text{.}
	 \label{17}%
\end{eqnarray}

We next analyze the laser state derived from Eq. (\ref{17}), by comparing it
with the squeezed vacuum state defined in Eq. (\ref{13}), which by its turn
comes from Eq. (\ref{15}). All the following figures consider the same
parameters used in Fig. 2 to validate the effective interaction in Eq.
(\ref{3}), together with the choices $\mathcal{C}/g=0.35$, $r/g=92$ and
$\gamma/g=0.5$, which lead to the rate $\mathcal{A}/g=736$.

In Fig. 4 we plot the mean occupation number $\left\langle a^{\dag
}a\right\rangle (t)$ of the laser state resulting from Eq. (\ref{17}). We
verify that the mean occupation number reaches the steady value $\left\langle
a^{\dag}a\right\rangle (t)=0.62$ for $gt\approx3$, long before the validity of
the effective Hamiltonian is compromised, for $gt\approx7$. This steady
excitation is in good agreement with that predicted by Eq. (\ref{13}), given
by $\left\langle a^{\dag}a\right\rangle =0.56$, about $10\%$ less than the
numerical simulation. The value $gt\approx3$ follows after the passage of
$1734$ atoms through the cavity.

In Fig. 5(a) we present the photon number distribution $\mathcal{P}_{n}$ and
the phase space projection of the Wigner function of the laser state following
from Eq. (\ref{17}), for $gt=4$. We again verify a good agreement with the
ideal squeezed vacuum state as it becomes clear from Fig. 5(b) where the
fidelity $\mathcal{F}=\operatorname*{Tr}\rho S(\xi)\left\vert 0\right\rangle
\left\langle 0\right\vert S^{\dag}(\xi)$ of the state $\rho$ coming from Eq.
(\ref{17}) is plotted against $gt$, regarded to the ideal squeezed vacuum
$S(\xi)\left\vert 0\right\rangle $ in Eq. (\ref{13}). We note that the
fidelity already starts from a high value due to the large population of the
 state in the squeezed vacuum field. The plot if Fig. 5(a) also
indicates that our laser has zero diffusion; if we had non-zero diffusion, the
elliptical projection should circulate around the origin of the phase space,
as occurs with the coherent state of the conventional laser theory \cite{M}.

In Figs. 6 we plot the variances $\left(  \Delta X_{1}\right)  ^{2}(t)$ (solid
line) and $\left(  \Delta X_{2}\right)  ^{2}(t)$ (dotted line), against $gt$,
of the laser state coming from Eq. (\ref{17}), showing that these variances
approaches the values $\left(  \Delta X_{1}\right)_{(S(\xi)\left\vert 0\right\rangle)}=\frac{e^{-2r}}{4}=0.06$ and $\left(  \Delta X_{2}\right)_{(S(\xi)\left\vert 0\right\rangle)}=\frac{e^{2r}}{4}=0.98$, computed from the squeezed vacuum
state in Eq. (\ref{13}). In Fig. 7 we plot the Mandel parameter $Q_{M}=\left(
\left\langle n^{2}\right\rangle -\left\langle n\right\rangle ^{2}\right)
/\left\langle n\right\rangle -1$ for the laser state, showing that for
$gt\approx2$ we reach a stationary value around $Q=2\cosh(\lvert r \rvert )-1=2.1,$ as expected for the
ideal squeezed vacuum state. Finally, in Fig. 8 we plot the off-diagonal
matrix elements $\rho_{0,8}(t)$ and $\rho_{4,6}(t)$, which demonstrate,
together with Fig. 5(a), that the phase coherence is preserved due to the
absence of phase diffusion of our squeezed vacuum laser, since $A\left\vert
\alpha\right\rangle _{A}$ $=0$; otherwise these matrix elements would decay to zero.

\section{Conclusion}

We have presented a method to produce a squeezed vacuum laser with zero
diffusion. This method is based on merging together the reservoir engineering
technique with the laser theory. The reservoir engineering demands us to build
up an effective interaction between the the system whose state we want to
protect (the cavity field) and an auxiliary system (the laser active medium),
of the form of Eq. (\ref{b}), $\chi\left(  A\sigma_{+}+A^{\dag}\sigma
_{-}\right)  $, with the laser steady state being an eigenstate of $A$ with
null eigenvalue: $A\left\vert \Psi\right\rangle =0$. The effective interaction
must\ enable the construction of an isomorphism between the field operators in
the effective $\left(  A^{\dag},A\right)  $ and the Jaynes-Cummings $\left(
a^{\dag},a\right)  $ Hamiltonians. The isomorphism is carried out by building
a basis state $\left\{  \left\vert n\right\rangle _{A}\right\}  $ for the
operators $A^{\dag},A$ similar to the Fock basis state $\left\{  \left\vert
n\right\rangle \right\}  $ for $a^{\dag},a$. The laser theory, by its turn,
provides the mechanism by which the cavity mode is fed by the stimulated
emission of the active medium when subjected to linear amplification, in
addition to inducing a saturation whichresults in a
far-from-equilibrium steady state.

Our method has an advantage over the reservoir engineering, in which we
cannot, evidently, eliminate the environment (described by the Lindbladian for
$a^{\dag},a$ in Eq. (\ref{R})) that acts to introduce errors in the action of
the artificially constructed reservoir (described by the Lindbladian for
$A^{\dag},A$). In our method, this does not occur, the only source of errors
being the protocol for building up the effective interaction, based on the
adiabatic elimination of fast variables, which is also present in reservoir
engineering. This is a very unique aspect of our method, which shows that the
association of reservoir engineering with the laser mechanism results in both
a more robust protocol than the reservoir engineering (since the Lindbladian
for $a^{\dag},a$ is absent) and also an unconventional laser field described
by a nonclassical coherence-preserving state.

In addition to the many applications we have already listed for a zero
linewidth laser, we observe that a squeezed vacuum laser may be useful for
high-resolution interferometry, a timely topic due to the newly emerging field
of gravitational-wave interferometry. Moreover, the present method
challenges us to design effective interactions leading to other nonclassical
laser states, as for example steady superposition states. This challenge can
leads us to a new chapter regarding the preparation of nonclassical steady
states, which could be useful for a variety of fundamental and technological applications.

\begin{flushleft}
{\Large \textbf{Acknowledgements}}
\end{flushleft}

FON and MHYM would like to thank CAPES and INCT-IQ for support.

\end{document}